\documentclass[twocolumn,english]{IEEEtran}
\usepackage[latin9]{inputenc}
\usepackage{geometry}
\usepackage{amsmath}
\usepackage{amssymb}
\usepackage{graphicx}



\usepackage{babel}
\begin{document}
\title{Stereographic coordinates for simpler far-field radiation analysis}
\author{\author{\IEEEauthorblockN{Vassili Savinov\\}
\IEEEauthorblockA{Independent Researcher \\ Southampton \\ United Kingdom \\ vsavinov@protonmail.com}}}

\maketitle
A coordinate system is proposed for the purposes of analysis of
far-field radiation from a localized source at the origin. The proposed
coordinates exhibit no singularities over the far-field hemi-sphere,
making them a good choice for automated radiation pattern analysis.
Finally, it is shown that Ludwig polarization basis, commonly used
to represent polarization of the far-field radiation, arises naturally,
as a coordinate basis, when one uses stereographic coordinates. 

\section{Introduction}

Evaluation of far-field radiation is a common task in antenna as well
as optical engineering. Distribution of radiation over the surface
of an imaginary sphere or hemi-sphere, with extremely large radius,
centered on the localized source of electromagnetic waves, is scrutinized
in terms of intensity, directivity, polarization state purity, coherence
as well as other key metrics \cite{BalanisBook}. 

To describe the distribution of the electromagnetic field on the far-field
sphere, it is natural to use spherical coordinate system as well as
spherical basis, for vector decomposition \cite{CollinBook1985}.
Given extraordinarily detailed theory of spherical coordinates \cite{FieldTheroyMoonSpencer,ArfkenWeberBook},
developed by generations of scientists and engineers, working with
them presents no serious difficulties for a skilled practitioner.
Getting computer algorithm to handle outputs in spherical coordinates,
however, can be a time-consuming exercise. The main difficulty stems
from handling the edge-cases presented by coordinate singularities
at south and north poles of the sphere, as well as the cyclic nature
of the azimuthal angle. Consider, for example, developing an algorithm
to compute and average value of electric field over a small, but finite
solid angle that could or could not include one of the poles. One
can certainly develop a robust and stable algorithm to handle such
cases, but there will be a price to pay in terms of human time spent
developing and maintaining the code. A better way to approach such
problems is to develop a suitable mathematical framework that removes
edge-cases by design.

Here we consider a mathematical solution to a slightly simpler, but
nevertheless common problem - analysis of far-field radiation on the
surface of a \emph{hemi}-sphere. The key difference is that unlike
the full sphere, the surface of a hemi-sphere can be described using
Cartesian-like coordinates, i.e. the surface of a hemi-sphere can
be mapped smoothly onto two-dimensional plane \cite{FlandersDiffForms}.
We develop key scalar and vector properties of one possible choice
for such coordinates, the stereographic coordinates. In the process,
we demonstrate that Ludwig basis \cite{Ludwig73}, a popular choice
for decomposing vector fields tangent to a sphere is essentially the
(normalized) coordinate basis for stereographic coordinates. Finally,
we present visualization and basic analysis of far-field radiation
from an illuminated aperture, in stereographic coordinates.

\section{Stereographic projection and Stereographic coordinates}

The roots of stereographic projection go back to the ancient Greece,
and in particular to work of Hipparchos. Whilst initial application
of stereographic projection was limited to astronomy, cartography
and navigation, its utility has since been recognized in crystallography
(e.g. Wulff net) and geology. A detailed historical account of stereographic
projection development and application has been complied by Howarth
\cite{Howarth1996}. Key mathematical properties of the projection
have been summarized by Rosenfeld and Sergeeva \cite{Rosenfeld1977}.
Despite the wide-spread use of stereographic projection in geo-sciences
\cite{LisleLeyshonBook,Haziot2018}, as well as engineering and material
science \cite{Rosenfeld1977,Howarth1996,Xiao2007,Liu2011,Rosca2009},
it remains relatively unknown in the field of applied electromagnetism.
In this section we provide two definitions of the stereographic projection:
geometric and algebraic.

\begin{figure}
\begin{centering}
\includegraphics{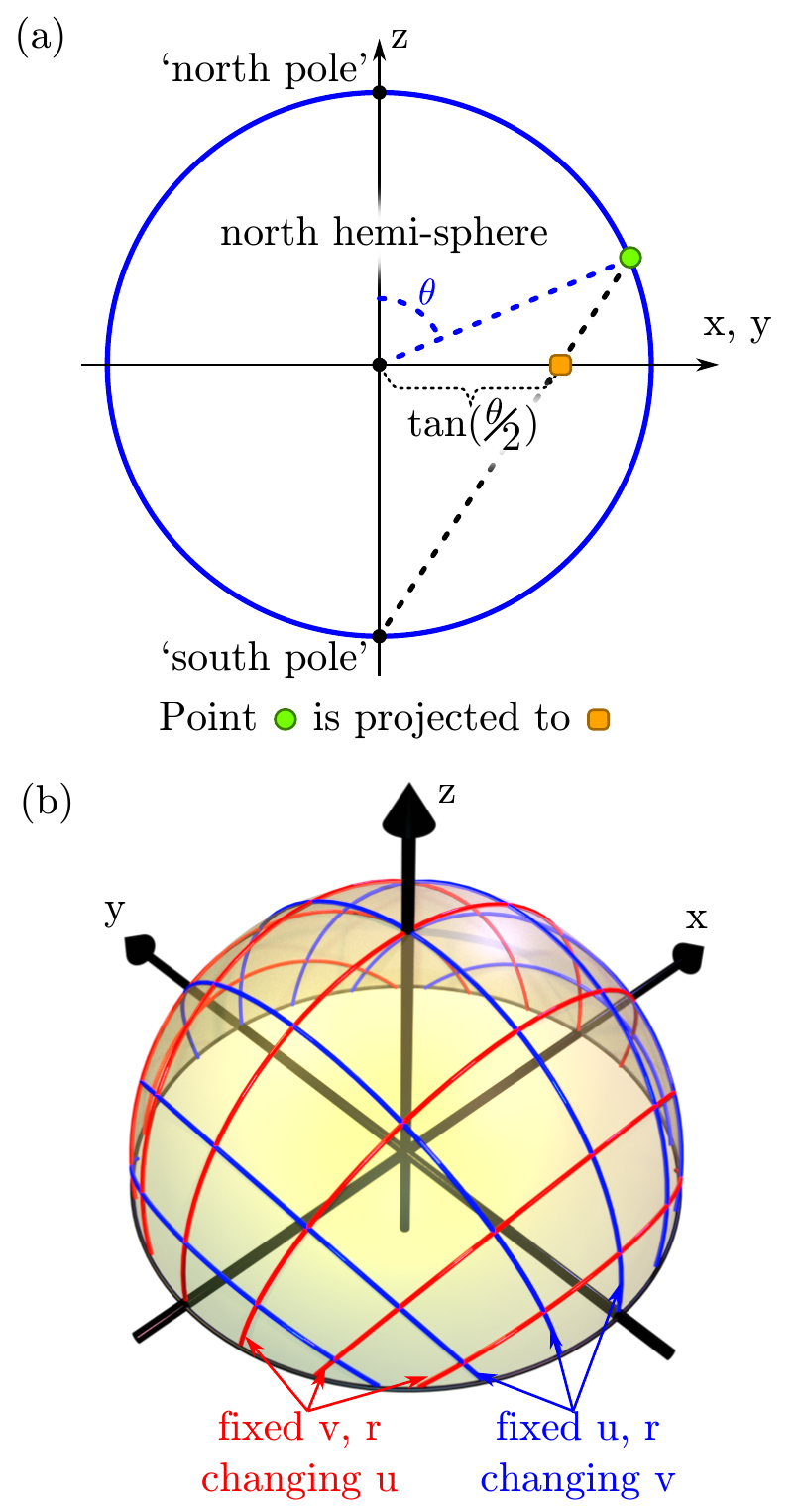}
\par\end{centering}
\caption{\textbf{Illustration of stereographic projection and stereographic
coordinates.} \textbf{(a)}~Geometric construction of the stereographic
projection. A point on a sphere is mapped onto a point on a 2d plane.
Here, for simplicity, we show a circle instead of the sphere and a
line instead of the plane. A chosen point on the circle (green circle),
which corresponds to angle $\theta$, is projected onto the horizontal
axis by plotting a straight line from the south pole of the circle
to the chosen point. The point at which the straight line intersects
the horizontal axis corresponds to the position of the projected point
(orange square). Formally, all points on the surface of a sphere,
parameterized by $\left(\theta,\,\phi\right)$, are projected onto
points on a plane with Cartesian coordinates $\left(\tan\frac{\theta}{2}\,\cos\phi,\,\tan\frac{\theta}{2}\,\sin\phi\right)$.
\textbf{(b)}~Grid generated by stereographic coordinates on a surface
of a hemi-sphere. One of the stereographic coordinates is the radius
$r$, which remains constant on the hemi-sphere. The other two coordinates
are $u$ and $v$. Red curves are generated by fixing $v$ (and $r$),
and varying $u$ within the limits set by $u^{2}+v^{2}=1$, blue curves,
correspondigly, are generated by fixing $u$ and varying $v$.}

\label{fig:stereographic_coord_illustration}
\end{figure}

The geometric definition of stereographic projection involves placing
a sphere, the surface of which is to be projected, with north- and
south-poles at $z$-axis, so that equator would lie in the $z=0$
plane, then drawing straight lines emanating from south-pole. The
point at which the straight line intersects the sphere is then mapped
onto point at which the same line intersects the $z=0$ plane. The
method is illustrated in Fig.~\ref{fig:stereographic_coord_illustration}a.
All of the sphere, except for the south-pole, can be mapped onto a
flat plane in this manner, however it is common to only map one hemi-sphere,
e.g. the north hemi-sphere, since mapped structures become progressively
more distorted as one comes closer to the south pole.

The algebraic definition of stereographic projection of an arbitrary
point $\left(r,\,\theta,\,\phi\right)$ in 3-dimensional space, in
spherical coordinates, to point $\text{\ensuremath{\left(\xi,\,\eta\right)}}$
in 2-dimensional space, in Cartesian coordinates, is:

\begin{flalign}
\xi= & \cos\phi\cdot\tan\frac{\theta}{2}\label{eq:RThPhitoXi}\\
\eta= & \sin\phi\cdot\tan\frac{\theta}{2}\label{eq:RThPhitoETa}
\end{flalign}

The projection remains well-defined for all points with $0\le\theta<\pi$.
Note that $\xi^{2}+\eta^{2}=\tan^{2}\theta/2$, thus all points, on
the sphere, with $\theta\le\pi/2$ are mapped inside the unit-radius
circle on the $\xi\eta$-plane. Thus stereographic projection, as
defined above, projects all points in the $z\ge0$ half-space into
a circle with unit radius in the $\xi\eta$-plane. 

\subsection{Stereographic coordinates}

Based on inverse stereographic projection, i.e. mapping unit-radius
circle to a surface of a hemi-sphere, one can define \emph{stereographic
coordinates for }$z\ge0$ half-space\emph{. }The stereographic coordinates
of an arbitrary point $P$ in $z\ge0$ half-space are $u$, $v$ and
$r$. The latter ($r$) is the distance from the origin to $P$, equivalently
it is the radius of the hemi-sphere that contains the point on its
surface. Coordinates $u$ and $v$ are the coordinates that point
$P$ would have if it were mapped onto a unit-circle with stereographic
projection. 

Figure~\ref{fig:stereographic_coord_illustration}b shows the curves
generated by fixing $r$ as well as either $u$ or $v$, and varying
the other. Crucially, the curves of varying $u$, i.e. the red curves
in Fig.~\ref{fig:stereographic_coord_illustration}b, do not intersect,
and the same applies to curves of varying $v$. This is a visual illustration
of an important property of the stereographic coordinates - lack of
the coordinate singularities. Stereographic coordinates allow one
to describe any point on a hemi-sphere in the same way as one would
describe any point inside a unit circle - with two Cartesian coordinates.
Coordinates $u$ and $v$ can vary within the range -1...1, and any
combination of these coordinates will by valid (only the points with
$u^{2}+v^{2}\le1$ will be on the northern hemi-sphere). Furthermore,
if two points on a hemi-sphere are close to each other, their $u$
and $v$ coordinates will always be close as well. Contrast this with,
spherical coordinates, where $\phi=0$ and $\phi=2\pi$ correspond
to the same location on a surface of the sphere, despite the large
difference in the value of the coordinate. This well-behavedness lends
itself readily for use in numerical analysis of, for example, far-field
radiation patterns, which will be discussed in Sec.~\ref{sec:Aperture}. 

The mapping between spherical coordinates and stereographic coordinates,
for $\theta\le\pi/2$, is given by ($r$ is common to both spherical
and stereographic coordinates):

\begin{flalign}
u= & \cos\phi\cdot\tan\frac{\theta}{2}\label{eq:SpherToU}\\
v= & \sin\phi\cdot\tan\frac{\theta}{2}\label{eq:SpherToV}\\
\theta= & 2\,\arctan\left(\sqrt{u^{2}+v^{2}}\right)\label{eq:UVtoTheta}\\
\phi= & \arctan\left(\frac{v}{u}\right)\label{eq:UVtoPhi}
\end{flalign}

Mapping between Cartesian and stereographic coordinates is:

\begin{flalign}
u= & \frac{x}{r+z}\label{eq:XYZtoU}\\
v= & \frac{y}{r+z}\label{eq:XYZtoV}\\
r= & \sqrt{x^{2}+y^{2}+z^{2}}\label{eq:XYZtoR}
\end{flalign}

\section{Stereographic basis and Ludwig basis}

One of the basic results in antenna theory is that far-field radiation
from a localized electromagnetic source is of transverse polarization
\cite{CollinBook1985}, i.e. tangent to the far-field sphere. Consequently,
far-field radiation can be expressed in spherical basis vectors $\boldsymbol{\hat{\theta}}$
(points from north to south pole, along the meridians) and $\boldsymbol{\hat{\phi}}$
(points west to east along parallels). However, use of spherical basis
becomes challenging at the poles of the hemi-sphere, where coordinate
singularity renders the spherical basis ill-defined. Ludwig polarization
basis \cite{Ludwig73,CollinBook1985} solves this problem by introducing
basis vectors:
\begin{flalign}
\boldsymbol{\hat{i}}_{ref}= & \sin\phi\,\boldsymbol{\hat{\theta}}+\cos\phi\,\boldsymbol{\hat{\phi}}\label{eq:LudwigRefSpher}\\
\boldsymbol{\hat{i}}_{cross}= & \cos\phi\,\boldsymbol{\hat{\theta}}-\sin\phi\,\boldsymbol{\hat{\phi}}\label{eq:LudwigCrossSpher}
\end{flalign}

Which are transverse, i.e. tangent to far-field sphere, orthonormal,
and well-behaved at poles. Here we will show that stereographic coordinates
naturally give rise to Ludwig basis.

Consider the coordinate basis of stereographic coordinates on the
surface of far-field hemi-sphere. The radial vector $\mathbf{\hat{r}}$
is the same as in spherical coordinates, and will therefore be skipped
in the following discussion. The coordinate basis vectors for $u$
and $v$, i.e. tangents to the curves of varying $u$ and $v$ (see
Fig.~\ref{fig:stereographic_coord_illustration}b), are: $\mathbf{e}_{u}=\frac{d}{du}$
and $\mathbf{e}_{v}=\frac{d}{dv}$ correspondingly. Here we follow
conventions of differential geometry in defining vectors as tangents
to curves \cite{SchutzMethods1980}. This is the coordinate basis
for the $uv$-coordinates on the far-field hemi-sphere. Using Eq.~(\ref{eq:SpherToU},
\ref{eq:SpherToV}), the relationship to the corresponding spherical
coordinate basis is then given by:
\begin{flalign*}
\text{\ensuremath{\mathbf{e}}}_{\theta}= & \frac{d}{d\theta}=\frac{\partial u}{\partial\theta}\mathbf{e}_{u}+\frac{\partial v}{\partial\theta}\mathbf{e}_{v}=\frac{\cos\phi}{2\cos^{2}\frac{\theta}{2}}\mathbf{e}_{u}+\frac{\sin\phi}{2\cos^{2}\frac{\theta}{2}}\mathbf{e}_{v}\\
= & \frac{1+u^{2}+v^{2}}{2\sqrt{u^{2}+v^{2}}}\,\left(u\mathbf{e}_{u}+v\mathbf{e}_{v}\right)\\
\text{\ensuremath{\mathbf{e}}}_{\phi}= & \frac{d}{d\phi}=\frac{\partial u}{\partial\phi}\mathbf{e}_{u}+\frac{\partial v}{\partial\phi}\mathbf{e}_{v}\\
= & -\tan\frac{\theta}{2}\,\sin\phi\,\mathbf{e}_{u}+\tan\frac{\theta}{2}\,\cos\phi\,\mathbf{e}_{v}\\
= & -v\mathbf{e}_{u}+u\mathbf{e}_{v}
\end{flalign*}

Converting to more familiar normalized basis for spherical coordinates:
$\boldsymbol{\hat{\theta}}=\mathbf{e}_{\theta}/r$, $\boldsymbol{\hat{\phi}}=\mathbf{e}_{\phi}/r\sin\theta$:
\begin{flalign*}
\boldsymbol{\hat{\theta}}= & \frac{1+u^{2}+v^{2}}{2r\sqrt{u^{2}+v^{2}}}\,\left(u\mathbf{e}_{u}+v\mathbf{e}_{v}\right)\\
\boldsymbol{\hat{\phi}}= & \frac{1+u^{2}+v^{2}}{2r\sqrt{u^{2}+v^{2}}}\,\left(-v\mathbf{e}_{u}+u\mathbf{e}_{v}\right)
\end{flalign*}

Similar approach, with Eq.~(\ref{eq:UVtoTheta},~\ref{eq:UVtoPhi}),
can be used to invert the relations:
\begin{flalign*}
\mathbf{e}_{u}= & \frac{2r}{\left(1+u^{2}+v^{2}\right)\sqrt{u^{2}+v^{2}}}\,\left(u\mathbf{\boldsymbol{\hat{\theta}}}-\,v\mathbf{\boldsymbol{\hat{\phi}}}\right)\\
\mathbf{e}_{v}= & \frac{2r}{\left(1+u^{2}+v^{2}\right)\sqrt{u^{2}+v^{2}}}\,\left(v\mathbf{\boldsymbol{\hat{\theta}}}+u\mathbf{\boldsymbol{\hat{\phi}}}\right)
\end{flalign*}

Clearly, $\mathbf{e}_{u}.\mathbf{e}_{v}=0$ and $\mathbf{e}_{u}.\mathbf{e}_{u}=\mathbf{e}_{v}.\mathbf{e}_{v}=4r^{2}/\left(1+u^{2}+v^{2}\right)^{2}$.
One can then define normalized stereographic basis as $\mathbf{\hat{u}}=\mathbf{e}_{u}\,\left(1+u^{2}+v^{2}\right)/2r$,~$\mathbf{\hat{v}}=\mathbf{e}_{v}\,\left(1+u^{2}+v^{2}\right)/2r$,
then:
\begin{flalign}
\boldsymbol{\hat{\theta}}= & \frac{1}{\sqrt{u^{2}+v^{2}}}\,\left(u\mathbf{\hat{u}}+v\mathbf{\hat{v}}\right)\label{eq:ThetaHatFromUV}\\
\boldsymbol{\hat{\phi}}= & \frac{1}{\sqrt{u^{2}+v^{2}}}\,\left(-v\mathbf{\hat{u}}+u\mathbf{\hat{v}}\right)\label{eq:PhiHatFromUV}
\end{flalign}

Finally, substituting Eq.~(\ref{eq:ThetaHatFromUV}, \ref{eq:PhiHatFromUV})
into Eq.~(\ref{eq:LudwigRefSpher}, \ref{eq:LudwigCrossSpher}) one
finds:

\begin{flalign}
\boldsymbol{\hat{i}}_{ref}= & \mathbf{\hat{v}}=\frac{1+u^{2}+v^{2}}{2r}\,\mathbf{e}_{v}\label{eq:LudwigRefUV}\\
\boldsymbol{\hat{i}}_{cross}= & \mathbf{\hat{u}}=\frac{1+u^{2}+v^{2}}{2r}\,\mathbf{e}_{u}\label{eq:LudwigCrossUV}
\end{flalign}

Ludwig basis is the normalized coordinate basis of the stereographic
coordinates.

One of the great utilities of coordinate basis is in providing a simple
form for common vector calculus operations. For example, for any suitably
smooth scalar field $f=f\left(u,\,v,\,r\right)$, its gradient is
given by:
\begin{flalign*}
\boldsymbol{\nabla}f= & \frac{\alpha}{2r}\,\frac{\partial f}{\partial u}\mathbf{\hat{u}}+\frac{\alpha}{2r}\,\frac{\partial f}{\partial v}\mathbf{\hat{v}}+\frac{\partial f}{\partial r}\,\mathbf{\hat{r}}\\
= & \frac{\alpha}{2r}\,\frac{\partial f}{\partial u}\boldsymbol{\hat{i}}_{cross}+\frac{\alpha}{2r}\,\frac{\partial f}{\partial u}\boldsymbol{\hat{i}}_{ref}+\frac{\partial f}{\partial r}\,\mathbf{\hat{r}}
\end{flalign*}

where $\alpha=1+u^{2}+v^{2}$ (to improve readability). See Supplementary
Materials (Sec.~\ref{app:vec_calc}) for further details, including
expressions for divergence and curl in stereographic coordinates.
Simplicity in extracting numerically well-behaved gradient, curl and
divergence can be of great utility in developing algorithms for identifying
and scrutinizing key metrics of far-field radiation from an antenna. 

\section{Application of stereographic coordinates: Visualization of far-field
radiation from uniformly illuminated aperture\label{sec:Aperture}}

As an application of stereographic coordinates, here we visualize
the far-field radiation from a square aperture that lies in the XY-plane,
at the origin. The size of the aperture is $a$, its sides are aligned
to lie parallel to $x$ and $y$ axis. The (time-harmonic) electric
field at the aperture follows:
\[
\mathbf{E}=\mathbf{\hat{x}}\cdot E_{0}\cdot\exp\left(-i\left(\zeta_{x}x+\zeta_{y}y\right)\right)\cdot\exp\left(i\omega t\right),\:-\frac{a}{2}\le x,y\le\frac{a}{2}
\]

and is zero at all other points on the $z=0$ plane. The phase gradient
was set to $\zeta_{x}=0.3k_{0}$ and $\zeta_{y}=0.6k_{0}$, where
$k_{0}=4\pi/a$. The angular frequency is $\omega=4\pi c/a$, where
$c$ is the speed of light, $t$ is time, and $E_{0}$ is the amplitude
of the electric field. The ambient environment is vacuum. The relevant
wavelength of the radiation at frequency $\omega$ is $\lambda_{0}=a/2$.

\begin{figure}
\begin{centering}
\includegraphics{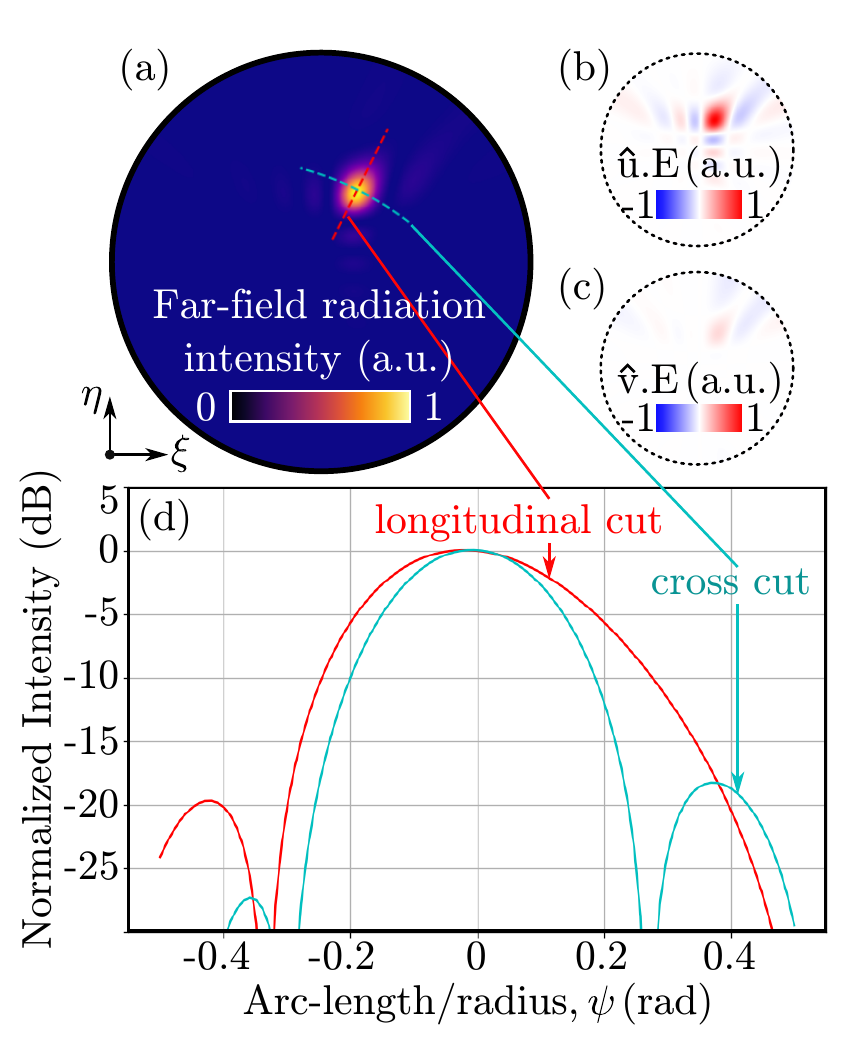}
\par\end{centering}
\caption{\textbf{Visualization of far-field radiation from square aperture
source. }The source electric field at the $z=0$ plane is $\mathbf{E}=\mathbf{\hat{x}}\cdot E_{0}\cdot\exp\left(-i\left(\zeta_{x}x+\zeta_{y}y\right)\right)\cdot\exp\left(i\omega t\right)$
inside the aperture and zero everywhere else. The phase gradient is
$\zeta_{x}=0.3k_{0}$ and $\zeta_{y}=0.6k_{0}$, where $k_{0}=\omega/c$
is the wavenumber of the radiation (and $c$ is the speed of light).
The wavelength of radiation is $\lambda_{0}=a/2$, where $a$ is the
size of the square aperture. \textbf{(a)}~Stereographic projection
of the intensity of the radiation, shown as colormap, emitted into
the north hemi-sphere. \textbf{(b),~(c)}~Show the colormaps of the
components of the far-field electric field along $\mathbf{\hat{u}}$
and $\mathbf{\hat{v}}$ respectively. \textbf{(d)}~Shows the cuts
through the intensity peak as a function of normalized arc-length
of the cut curve. Since the arc-length is normalized with respect
to the radius of the hemi-sphere, its units are radians. The cuts
are also depicted in (a).}

\label{fig:illum_aperture}
\end{figure}

The full expression for the far-field radiation from the source described
above, in stereographic coordinates, is given in the Appendix (see
App.~\ref{app:illum_app}). Since the expression for the radiation
pattern, in spherical coordinates, is well-known \cite{CollinBook1985},
and its conversion to stereographic coordinates is a straightforward
algebraic exercise, here we focus only on the visualization of the
radiation pattern. Figure~\ref{fig:illum_aperture}a, shows the intensity
of the far-field radiation over the $z\ge0$ hemi-sphere, projected
into a circle via the stereographic projection. To avoid confusion,
we define $\xi,\,\eta$ to be the Cartesian coordinates on the plane
used for visualization, i.e. the plane of Fig.~\ref{fig:illum_aperture}a.
The stereographic projection of a point at $\left(u,\,v,\,r\right)$,
in 3d stereographic coordinates, onto the visualization plane can
then be expressed as $\xi=u$ and $\eta=v$. Figures~\ref{fig:illum_aperture}b,c
show the $\mathbf{\hat{u}}=\boldsymbol{\hat{i}}_{cross}$ and $\mathbf{\hat{v}}=\boldsymbol{\hat{i}}_{ref}$
components of the electric field, respectively. The nature of stereographic
projection means that $\mathbf{\hat{u}}$-polarized fields in stereographic
coordinates will be pointing horizontally, i.e. along $\boldsymbol{\hat{\xi}}$,
on the $\xi\eta$-plane, after projection. Correspondingly, the $\mathbf{\hat{v}}$-polarized
fields will be projected into the vertically-pointed fields on $\xi\eta$-plane.

Figure~\ref{fig:illum_aperture}d shows the meridional and cross
cuts through the peak of the radiation intensity. The same cuts are
also indicated in Fig.~\ref{fig:illum_aperture}a. Both meridional
and cross cuts are along the so-called great circles \cite{SchutzMethods1980},
i.e. the curves that correspond to `straight' lines on a spherical
surface (geodesics). An easy way to visualize these curves is to note
that \emph{if} the peak of the radiation intensity was at the north-pole
of the far-field hemi-sphere, then the meridional cut could be along
the $0^{\circ}$ and $180^{\circ}$ meridian, whilst the cross cut
would be along $90^{\circ}$ and $270^{\circ}$ meridian. Expressions
for arbitrary great-circle curves, in stereographic coordinates, are
given in the Supplementary Material (see Sec.~\ref{app:great_circles}). 

Apart from cuts through the peak of radiation intensity it may be
necessary to calculate the power radiated into a certain solid angle.
Such calculations need to take into account the Jacobian of the stereographic
coordinates $J=\left(2r/\left(1+u^{2}+v^{2}\right)\right)^{2}$ (see
Supplementary Materials; Sec.~\ref{app:vec_calc}). Thus, given radiant
intensity, i.e. power per solid angle, $I_{\Omega}\left(u,\,v\right)\propto r^{2}\left|\mathbf{E}\left(u,\,v,\,r\right)\right|^{2}$,
where $\mathbf{E}$ is the electric field. The power ($P_{\Xi}$)
radiated into a specific solid angle $\Xi$, i.e. a region in Fig.~\ref{fig:illum_aperture}a,
can be calculated as:
\[
P_{\Xi}=\int_{\Xi}du\,dv\,I_{\Omega}\left(u,\,v\right)\,\left(\frac{2}{1+u^{2}+v^{2}}\right)^{2}
\]

\section{Conclusion}

In summary, we have developed key properties of the stereographic
coordinate system, which is induced by the (inverse) stereographic
projection. Apart from providing a convenient, singularity-free, way
of describing scalar fields on a surface of a hemi-sphere, such as
radiation pattern from an illuminated aperture, for example, the stereographic
coordinates give rise to singularity-free coordinate basis. Here we
have shown that the stereographic coordinate basis is the Ludwig basis,
already well-known in the field of antenna design. Knowledge of the
coordinate system that gives rise to Ludwig basis, allows one, for
example, to develop vector calculus tailored to Ludwig basis. Looking
ahead, we expect that stereographic coordinates will find wide-spread
use in the field of applied electromagnetism, particularly in areas
where computers need to be taught to analyze radiation patterns from
antennas, such as AI-aided antenna synthesis. 

\bibliographystyle{unsrt}

\begin{thebibliography}{10}

\bibitem{BalanisBook}
C.~A. Balanis.
\newblock {\em Antenna theory}.
\newblock John Wiley \& Sons, 3rd edition, 2005.

\bibitem{CollinBook1985}
C.~E. Collin.
\newblock {\em Antennas and Radiowave Propagation}.
\newblock McGraw-Hill, 1985.

\bibitem{FieldTheroyMoonSpencer}
P.~Moon and D.~E. Spencer.
\newblock {\em Field theory handbook}.
\newblock Springer-Verlag, 2nd edition, 1971.

\bibitem{ArfkenWeberBook}
{G. B. Arfken, H. J. Weber, and F. E. Harris}.
\newblock {\em Mathematical methods for physicists}.
\newblock Academic Press, 7th edition, 2013.

\bibitem{FlandersDiffForms}
H.~Flanders.
\newblock {\em Differential Forms}.
\newblock Dover Publications, 1989.

\bibitem{Ludwig73}
A.~C. Ludwig.
\newblock The definition of cross polarization.
\newblock {\em IEEE Tran. Antennas Propag.}, AP-21:116--119, 1973.

\bibitem{Howarth1996}
R.~J. Howarth.
\newblock History of the stereographic projection and its early use in geology.
\newblock {\em Terra Review}, page 499, 1996.

\bibitem{Rosenfeld1977}
{B. A. Rosenfeld and N. D. Sergeeva}.
\newblock {\em Stereographic Projection}.
\newblock MIR Publishers, Moscow, 1977.

\bibitem{LisleLeyshonBook}
R.~J. Lisle and P.R. Leyshon.
\newblock {\em Stereographic projection techniques for geologists and civil
  engineers}.
\newblock Cambridge University Press, 2nd edition, 2014.

\bibitem{Haziot2018}
{S. V. Haziot, and K. Marynets}.
\newblock Applying the stereographic projection to modeling of the flow of the
  antarctic circumpolar current.
\newblock {\em Oceanography}, 31(3):68--75, 2018.
\newblock https://doi.org/10.5670/oceanog.2018.311.

\bibitem{Xiao2007}
C.~Xiao and M.~G. Rossmann.
\newblock Interpretation of electron density with stereographic roadmap
  prjections.
\newblock {\em Journal of Structural Biology}, 158:182--187, 2007.

\bibitem{Liu2011}
H.~Liu and J.~Liu.
\newblock Sp2: a computer program for plotting stereographic projection and
  exploring crystallographic orientation relationships.
\newblock {\em Journal of Applied Crystallography}, 45:130--134, 2011.

\bibitem{Rosca2009}
D.~Ro{\c s}ca and J.-P. Antoine.
\newblock Locally supported orthogonal wavelet bases on the sphere via
  stereographic projection.
\newblock {\em Mathematical Problems in Engineering}, 2009:124904, 2009.
\newblock doi:10.1155/2009/124904.

\bibitem{SchutzMethods1980}
B.~Schutz.
\newblock {\em Geometrical methods of mathematical physics}.
\newblock Cambridge University Press, 1999.

\bibitem{LovelockRund}
D.~Lovelock and H.~Rund.
\newblock {\em Tensors, differential forms, and variational principles}.
\newblock Dover Publications, 1989.

\bibitem{Bachman06}
D.~Bachman.
\newblock {\em A geometric approach to differential forms}.
\newblock Birkh{\"a}user, 2nd edition, 2006.

\bibitem{LeeManifolds2013}
J.~M. Lee.
\newblock {\em Introduction to smooth manifolds}.
\newblock Springer, New York Heidelberg Dordrecht London, 2013.

\bibitem{FrankelGeomBook}
T.~Frankel.
\newblock {\em The geometry of physics}.
\newblock Cambridge University Press, 3rd edition, 2012.

\end{thebibliography}

\newpage{}

\section{Appendix: Vector calculus\label{app:vec_calc}}

The aim of this section is to calculate the common vector-calculus
expressions in stereographic coordinates. To accomplish this task
we will be relying on exterior derivative $\mathbf{d}\dots$, Hodge
star $\star$, as well as normal $\mathbf{g}$ and and inverse $\mathbf{g^{-1}}$
metric tensors \cite{SchutzMethods1980}. Starting with the latter
two, metric tensor is a bi-linear map that maps two vectors onto a
real-valued scalar (at every point of the considered manifold). Representing
the metric tensor as a matrix, for the case of stereographic coordinates:
\begin{flalign*}
\mathbf{g}= & \left(\begin{array}{ccc}
g_{uu} & 0 & 0\\
0 & g_{vv} & 0\\
0 & 0 & g_{rr}
\end{array}\right)\\
= & \left(\begin{array}{ccc}
\mathbf{e}_{u}.\mathbf{e}_{u} & 0 & 0\\
0 & \mathbf{e}_{v}.\mathbf{e}_{v} & 0\\
0 & 0 & \mathbf{\hat{r}}.\mathbf{\hat{r}}
\end{array}\right)\\
= & \left(\begin{array}{ccc}
\frac{4r^{2}}{\alpha^{2}} & 0 & 0\\
0 & \frac{4r^{2}}{\alpha^{2}} & 0\\
0 & 0 & 1
\end{array}\right)
\end{flalign*}

Where $\alpha=1+u^{2}+v^{2}$. The expression above can be used to
extract the Jacobian for the stereographic coordinates \cite{LovelockRund}:
\[
J=\sqrt{\mbox{det}\left[\mathbf{g}\right]}=\frac{4r^{2}}{\alpha^{2}}
\]

We denote the map from a single vector to a real-valued scalar, based
on some vector $\mathbf{q}$, as $\mathbf{g}\left(\mathbf{q},\,\bullet\right)$.
Then for any vector $\mathbb{\mathbf{p}}$:
\[
\left(\mathbf{g}\left(\mathbf{q},\,\bullet\right)\right)\left(\mathbf{p}\right)=\mathbf{g}\left(\mathbf{q},\,\mathbf{p}\right)=scalar
\]

Since $\mathbf{g}\left(\mathbf{q},\,\bullet\right)$ maps vectors
to scalars (and is linear), it is a 1-form \cite{Bachman06}. The
metric tensor maps vectors to 1-forms (this is also known as musical
isomorphisms \cite{LeeManifolds2013}). Using non-normalized basis
for vectors (and bearing in mind that $\mathbf{e}_{r}=\frac{d}{dr}=\mathbf{\hat{r}}$,
i.e. radial basis vector is normalized from the outset):
\begin{flalign*}
\mathbf{g}\left(\mathbf{e}_{u},\,\bullet\right)= & \frac{4r^{2}}{\alpha^{2}}\,\mathbf{d}u\\
\mathbf{g}\left(\mathbf{e}_{v},\,\bullet\right)= & \frac{4r^{2}}{\alpha^{2}}\,\mathbf{d}v\\
\mathbf{g}\left(\mathbf{\hat{r}},\,\bullet\right)= & \mathbf{d}r
\end{flalign*}

The inverse metric tensor maps pairs of 1-forms to real-valued scalars:
\begin{flalign*}
\mathbf{g}^{-1} & =\left(\begin{array}{ccc}
g^{uu} & 0 & 0\\
0 & g^{vv} & 0\\
0 & 0 & g^{rr}
\end{array}\right)\\
= & \left(\begin{array}{ccc}
\mathbf{d}u.\mathbf{d}u & 0 & 0\\
0 & \mathbf{d}v.\mathbf{d}v & 0\\
0 & 0 & \mathbf{d}r.\mathbf{d}r
\end{array}\right)\\
= & \left(\begin{array}{ccc}
1/g_{uu} & 0 & 0\\
0 & 1/g_{vv} & 0\\
0 & 0 & 1/g_{rr}
\end{array}\right)\\
= & \left(\begin{array}{ccc}
\frac{\alpha^{2}}{4r^{2}} & 0 & 0\\
0 & \frac{\alpha^{2}}{4r^{2}} & 0\\
0 & 0 & 1
\end{array}\right)
\end{flalign*}

Equivalently:
\begin{flalign*}
\mathbf{g}^{-1}\left(\mathbf{d}u,\,\bullet\right)= & \frac{\alpha^{2}}{4r^{2}}\,\mathbf{e}_{u}\\
\mathbf{g}^{-1}\left(\mathbf{d}v,\,\bullet\right)= & \frac{\alpha^{2}}{4r^{2}}\,\mathbf{e}_{v}\\
\mathbf{g}^{-1}\left(\mathbf{d}r,\,\bullet\right)= & \mathbf{\hat{r}}
\end{flalign*}

The volume form in stereographic coordinates is $\boldsymbol{\omega}=\sqrt{g}\mathbf{d}u\wedge\mathbf{d}v\wedge\mathbf{d}r=\frac{4r^{2}}{\alpha^{2}}\mathbf{d}u\wedge\mathbf{d}v\wedge\mathbf{d}r$
\cite{LovelockRund,FrankelGeomBook}. Based on it the action of Hodge
star on individual forms is:
\begin{flalign*}
\star1= & \frac{4r^{2}}{\alpha^{2}}\mathbf{d}u\wedge\mathbf{d}v\wedge\mathbf{d}r\\
\star\mathbf{d}u= & \mathbf{d}v\wedge\mathbf{d}r\\
\star\mathbf{d}v= & \mathbf{d}r\wedge\mathbf{d}r\\
\star\mathbf{d}r= & \frac{4r^{2}}{\alpha^{2}}\,\mathbf{d}u\wedge\mathbf{d}v\\
\star\left(\mathbf{d}u\wedge\mathbf{d}v\right)= & \frac{\alpha^{2}}{4r^{2}}\mathbf{d}r\\
\star\left(\mathbf{d}v\wedge\mathbf{d}r\right)= & \mathbf{d}u\\
\star\left(\mathbf{d}r\wedge\mathbf{d}u\right)= & \mathbf{d}v\\
\star\left(\mathbf{d}u\wedge\mathbf{d}v\wedge\mathbf{d}r\right)= & \frac{\alpha^{2}}{4r^{2}}
\end{flalign*}

Using the equations above one can readily tackle a large portion of
the standard tasks in vector calculus, as will be shown in the following
sub-sections.

\subsection{Gradient}

The equation for vector-field $\boldsymbol{\nabla}f$, i.e. a gradient
of some function, is \cite{FlandersDiffForms,FrankelGeomBook}:
\begin{flalign*}
\boldsymbol{\nabla}f= & g^{-1}\left(\mathbf{d}f,\,\bullet\right)\\
= & g^{-1}\left(\partial_{u}f\mathbf{d}u+\partial_{v}f\mathbf{d}v+\partial_{r}f\mathbf{d}r,\,\bullet\right)\\
= & \frac{\alpha^{2}}{4r^{2}}\partial_{u}f\mathbf{e}_{u}+\frac{\alpha^{2}}{4r^{2}}\partial_{v}f\mathbf{e}_{v}+\partial_{r}f\mathbf{\hat{r}}
\end{flalign*}

Using Eq.~(\ref{eq:LudwigRefUV},~\ref{eq:LudwigCrossUV}), i.e.
$\mathbf{e}_{u}=\boldsymbol{\hat{i}}_{cross}\cdot2r/\alpha$ and $\mathbf{e}_{v}=\boldsymbol{\hat{i}}_{ref}\cdot2r/\alpha$:
\begin{flalign*}
\boldsymbol{\nabla}f= & \frac{\alpha}{2r}\partial_{u}f\mathbf{\hat{u}}+\frac{\alpha}{2r}\partial_{v}f\mathbf{\hat{v}}+\partial_{r}f\mathbf{\hat{r}}\\
= & \frac{\alpha}{2r}\partial_{u}f\boldsymbol{\hat{i}}_{cross}+\frac{\alpha}{2r}\partial_{v}f\boldsymbol{\hat{i}}_{ref}+\partial_{r}f\mathbf{\hat{r}}
\end{flalign*}

\subsection{Divergence}

Given vector field $\mathbf{A}=A^{u}\mathbf{e}_{u}+A^{v}\mathbf{e}_{v}+A^{r}\mathbf{e}_{r}$,
where $\mathbf{e}_{r}=\mathbf{\hat{r}}$, the equation for its divergence
becomes \cite{FlandersDiffForms,FrankelGeomBook}:
\begin{flalign*}
\boldsymbol{\nabla}.\mathbf{A}= & \star\mathbf{d}\star\mathbf{g}\left(\mathbf{A},\,\bullet\right)\\
= & \star\mathbf{d}\star\left(\mathbf{d}u\,\frac{4r^{2}}{\alpha^{2}}A^{u}+\mathbf{d}v\,\frac{4r^{2}}{\alpha^{2}}A^{v}+\mathbf{d}r\,A^{r}\right)\\
= & \star\mathbf{d}\left(\frac{4r^{2}}{\alpha^{2}}A^{u}\mathbf{d}v\wedge\mathbf{d}r\right)+\star\mathbf{d}\left(\frac{4r^{2}}{\alpha^{2}}A^{v}\mathbf{d}v\wedge\mathbf{d}r\right)+\\
 & +\star\mathbf{d}\left(\frac{4r^{2}}{\alpha^{2}}A^{r}\mathbf{d}u\wedge\mathbf{d}v\right)\\
= & \star\Biggl\{\Biggl(\partial_{u}\left(\frac{4r^{2}}{\alpha^{2}}A^{u}\right)+\partial_{v}\left(\frac{4r^{2}}{\alpha^{2}}A^{v}\right)+\\
 & +\partial_{r}\left(\frac{4r^{2}}{\alpha^{2}}A^{r}\right)\Biggr)\mathbf{d}u\wedge\mathbf{d}v\wedge\mathbf{d}r\Biggr\}\\
= & \frac{\alpha^{2}}{4r^{2}}\Biggl\{\Biggl(\partial_{u}\left(\frac{4r^{2}}{\alpha^{2}}A^{u}\right)+\partial_{v}\left(\frac{4r^{2}}{\alpha^{2}}A^{v}\right)+\\
 & +\partial_{r}\left(\frac{4r^{2}}{\alpha^{2}}A^{r}\right)\Biggr)\Biggr\}\\
= & \alpha^{2}\,\partial_{u}\left(\frac{A^{u}}{\alpha^{2}}\right)+\alpha^{2}\,\partial_{v}\left(\frac{A^{v}}{\alpha^{2}}\right)+\frac{1}{r^{2}}\,\partial_{r}\left(r^{2}A^{r}\right)
\end{flalign*}

Using Eq.~(\ref{eq:LudwigRefUV},~\ref{eq:LudwigCrossUV}), i.e.
$\mathbf{e}_{u}=\boldsymbol{\hat{i}}_{cross}\cdot2r/\alpha$ and $\mathbf{e}_{v}=\boldsymbol{\hat{i}}_{ref}\cdot2r/\alpha$:
\begin{flalign*}
\mathbf{A}= & A^{u}\mathbf{e}_{u}+A^{v}\mathbf{e}_{v}+A^{r}\mathbf{\hat{r}}\\
= & \frac{2r}{\alpha}A^{u}\boldsymbol{\hat{i}}_{cross}+\frac{2r}{\alpha}A^{u}\boldsymbol{\hat{i}}_{ref}+A^{r}\mathbf{\hat{r}}
\end{flalign*}

Thus:
\begin{flalign*}
A^{u}= & \frac{\alpha}{2r}\boldsymbol{\hat{i}}_{cross}.\mathbf{A}=\frac{\alpha}{2r}\mathbf{\hat{u}}.\mathbf{A}\\
A^{v}= & \frac{\alpha}{2r}\boldsymbol{\hat{i}}_{ref}.\mathbf{A}=\frac{\alpha}{2r}\mathbf{\hat{v}}.\mathbf{A}\\
A^{r}= & \mathbf{\hat{r}}.\mathbf{A}
\end{flalign*}

Therefore:
\begin{flalign*}
\boldsymbol{\nabla}.\mathbf{A}= & \alpha^{2}\,\partial_{u}\left(\frac{\mathbf{\hat{u}}.\mathbf{A}}{2r\alpha}\right)+\alpha^{2}\,\partial_{v}\left(\frac{\mathbf{\hat{v}}.\mathbf{A}}{2r\alpha}\right)+\\
 & +\frac{1}{r^{2}}\,\partial_{r}\left(r^{2}\mathbf{\hat{r}}.\mathbf{A}\right)
\end{flalign*}

\subsection{Curl}

Curl of a vector field $\mathbf{A}=A^{u}\mathbf{e}_{u}+A^{v}\mathbf{e}_{v}+A^{r}\mathbf{e}_{r}$,
where $\mathbf{e}_{r}=\mathbf{\hat{r}}$, can be expressed as \cite{FlandersDiffForms,FrankelGeomBook}:
\[
\boldsymbol{\nabla}\times\mathbf{A}=\mathbf{g}^{-1}\left(\star\mathbf{d}\mathbf{g}\left(\mathbf{A},\,\bullet\right),\,\bullet\right)
\]

Ignoring the last step for now:
\begin{flalign*}
\star\mathbf{d}\mathbf{g}\left(\mathbf{A},\,\bullet\right)= & \star\mathbf{d}\left(\mathbf{d}u\,\frac{4r^{2}}{\alpha^{2}}A^{u}+\mathbf{d}v\,\frac{4r^{2}}{\alpha^{2}}A^{v}+\mathbf{d}r\,A^{r}\right)\\
= & \star\biggl(-\mathbf{d}u\wedge\mathbf{d}v\,\partial_{v}\left(\frac{4r^{2}}{\alpha^{2}}A^{u}\right)+\\
 & +\mathbf{d}r\wedge\mathbf{d}u\,\partial_{r}\left(\frac{4r^{2}}{\alpha^{2}}A^{u}\right)+\\
 & +\mathbf{d}u\wedge\mathbf{d}v\,\partial_{u}\left(\frac{4r^{2}}{\alpha^{2}}A^{v}\right)-\\
 & -\mathbf{d}v\wedge\mathbf{d}r\,\partial_{r}\left(\frac{4r^{2}}{\alpha^{2}}A^{v}\right)-\\
 & \,\,-\mathbf{d}r\wedge\mathbf{d}u\,\partial_{u}\left(A^{r}\right)+\\
 & +\mathbf{d}v\wedge\mathbf{d}r\,\partial_{v}\left(A^{r}\right)\biggr)\\
= & \,\,\left(\partial_{v}\left(A^{r}\right)-\partial_{r}\left(\frac{4r^{2}}{\alpha^{2}}A^{v}\right)\right)\cdot\mathbf{d}u+\\
 & +\left(\partial_{r}\left(\frac{4r^{2}}{\alpha^{2}}A^{u}\right)-\partial_{u}\left(A^{r}\right)\right)\cdot\mathbf{d}v+\\
 & +\biggl(\partial_{u}\left(\frac{4r^{2}}{\alpha^{2}}A^{v}\right)-\\
 & -\partial_{v}\left(\frac{4r^{2}}{\alpha^{2}}A^{u}\right)\biggr)\cdot\frac{\alpha^{2}}{4r^{2}}\cdot\mathbf{d}r
\end{flalign*}

Applying the last step to change 1-forms to vectors: 
\begin{multline*}
\mathbf{g}^{-1}\left(\star\mathbf{d}\mathbf{g}\left(\mathbf{A},\,\bullet\right),\,\bullet\right)=\\
=\left(\partial_{v}\left(A^{r}\right)-\partial_{r}\left(\frac{4r^{2}}{\alpha^{2}}A^{v}\right)\right)\cdot\frac{\alpha^{2}}{4r^{2}}\mathbf{e}_{u}+\\
+\left(\partial_{r}\left(\frac{4r^{2}}{\alpha^{2}}A^{u}\right)-\partial_{u}\left(A^{r}\right)\right)\cdot\frac{\alpha^{2}}{4r^{2}}\mathbf{e}_{u}+\\
+\left(\partial_{u}\left(\frac{4r^{2}}{\alpha^{2}}A^{v}\right)-\partial_{v}\left(\frac{4r^{2}}{\alpha^{2}}A^{u}\right)\right)\cdot\frac{\alpha^{2}}{4r^{2}}\cdot\mathbf{\hat{r}}
\end{multline*}

Finally, expressing all in normalized basis:
\begin{flalign*}
\boldsymbol{\nabla}\times\mathbf{A}= & \,\,\frac{\alpha}{2r}\cdot\left(\partial_{v}\left(\mathbf{\hat{r}}.\mathbf{A}\right)-\partial_{r}\left(\frac{2r}{\alpha}\mathbf{\hat{v}}.\mathbf{A}\right)\right)\cdot\mathbf{\hat{u}}+\\
 & +\frac{\alpha}{2r}\cdot\left(\partial_{r}\left(\frac{2r}{\alpha}\mathbf{\hat{u}}.\mathbf{A}\right)-\partial_{u}\left(\mathbf{\hat{r}}.\mathbf{A}\right)\right)\cdot\mathbf{\hat{v}}+\\
 & +\frac{\alpha^{2}}{4r^{2}}\cdot\left(\partial_{u}\left(\frac{2r}{\alpha}\mathbf{\hat{v}}.\mathbf{A}\right)-\partial_{v}\left(\frac{2r}{\alpha}\mathbf{\hat{u}}.\mathbf{A}\right)\right)\cdot\mathbf{\hat{r}}
\end{flalign*}

\section{Appendix: Great circles on the hemi-sphere\label{app:great_circles}}

The aim of this section is to derive equation for an arbitrary great
circle, i.e. a curve on the surface of a sphere created by intersection
of the sphere surface with a plane that passes through the centre
of the sphere. Great circles are geodesic curves on spheres \cite{SchutzMethods1980}.
For purposes of this section it is easier to derive main results in
Cartesian coordinates $xyz$ and then to connect them to stereographic
coordinates using Eq.~(\ref{eq:XYZtoU},~\ref{eq:XYZtoV},~\ref{eq:XYZtoR}).

The great circle will be obtained by starting from a great circle
in the xz-plane that goes around the y-axis in counter-clockwise direction:
\[
\boldsymbol{\mathcal{C}}'''\left(\psi\right)=\left(\begin{array}{c}
r\sin\psi\\
0\\
r\cos\psi
\end{array}\right)
\]

The aim is to rotate this great circle to go through point $P_{0}=\left(\theta_{0},\,\phi_{0}\right)=\left(60^{\circ},\,30^{\circ}\right)$,
in spherical coordinates, at $\psi=0$. Simultaneously it is desirable
to have an additional parameter to control in which direction the
great circle points as $\psi$ grows, call this parameter $\Phi$.
Begin by rotating the curve $\boldsymbol{\mathcal{C}}'''$ around
the $z$ by $\Phi$ to get $\boldsymbol{\mathcal{C}}'$':
\begin{flalign*}
\boldsymbol{\mathcal{C}}''\left(\psi\right)= & \left(\begin{array}{ccc}
\cos\Phi & -\sin\Phi & 0\\
\sin\Phi & \cos\Phi & 0\\
0 & 0 & 1
\end{array}\right)\boldsymbol{\mathcal{C}}'''\left(\psi\right)\\
= & \left(\begin{array}{c}
r\cos\Phi\,\sin\psi\\
r\sin\Phi\,\sin\psi\\
r\cos\psi
\end{array}\right)
\end{flalign*}

The result, $\boldsymbol{\mathcal{C}}''$, still goes through north
pole , but the tangent to $\boldsymbol{\mathcal{C}}''$, at the north
pole, is now at angle $\Phi$ relative to the y-axis. Next rotate
$\boldsymbol{\mathcal{C}}''$ around y-axis by angle $\theta_{0}$
to get $\boldsymbol{\mathcal{C}}'$:

\begin{flalign*}
\boldsymbol{\mathcal{C}}'\left(\psi\right)= & \left(\begin{array}{ccc}
\cos\theta_{0} & 0 & \sin\theta_{0}\\
0 & 1 & 0\\
-\sin\theta_{0} & 0 & \cos\theta_{0}
\end{array}\right)\boldsymbol{\mathcal{C}}''\left(\psi\right)\\
= & \left(\begin{array}{c}
r\cos\Phi\,\cos\theta_{0}\,\sin\psi+r\sin\theta_{0}\,\cos\psi\\
r\sin\left(\Phi\right)\,\sin\psi\\
-r\cos\Phi\,\sin\theta_{0}\,\sin\psi+r\cos\theta_{0}\,\cos\psi
\end{array}\right)
\end{flalign*}

This will bring the point of $\psi=0$ to the correct latitude ($\theta_{0}$)
of the destination. Finally, rotate around z-axis by $\phi_{0}$ to
get curve $\boldsymbol{\mathcal{C}}$ which has point $\psi=0$ at
$P_{0}$:

\begin{multline*}
\boldsymbol{\mathcal{C}}\left(\psi\right)=\left(\begin{array}{ccc}
\cos\left(\phi_{0}\right) & -\sin\left(\phi_{0}\right) & 0\\
\sin\left(\phi_{0}\right) & \cos\left(\phi_{0}\right) & 0\\
0 & 0 & 1
\end{array}\right)\boldsymbol{\mathcal{C}}'\left(\psi\right)
\end{multline*}

\begin{figure}
\begin{centering}
\includegraphics{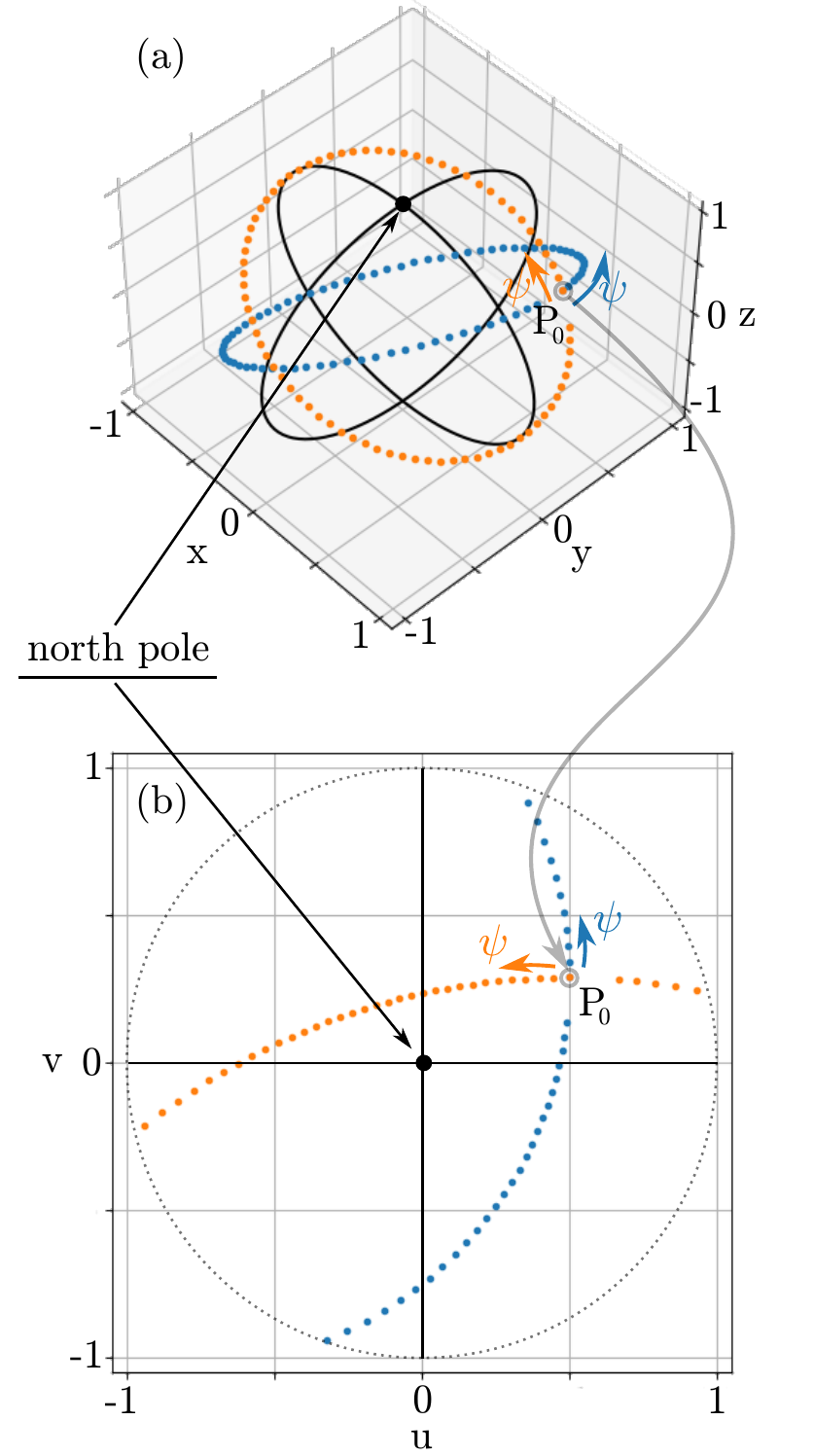}\caption{\textbf{Two great circles intersecting at a common point $P_{0}$}.
\textbf{(a)}~The actual shape of the two great circles in the 3d
space. The great circles, orange and blue, are plotted as sets of
disconnected dots distributed uniformly (fixed steps in $\psi$) along
the great circle. Parameter $\psi$ is the arc-length along the great
circles divided by the radius of the sphere. Point $P_{0}$ corresponds
to $\psi=0$ on both great circles. The orange and blue arrows point
in the direction of growing $\psi$. The last section of the blue
and orange great circles, i.e. where $\psi$ is at maximum value (see
Eq.~(\ref{eq:great_circle_u},~\ref{eq:great_circle_v})), is omitted
to show where each great circle starts and ends. The black circles
are meridians used to outline the underlying sphere. Meridian intersection
at the north pole is marked with a bold dot. \textbf{(b)}~The $u$
and $v$ stereographic coordinates for the points of the two great
circles in the north hemi-sphere. The changing distance between the
dots in the projected great circles shows the distortion that arises
as a result of stereographic projection.}
\par\end{centering}
\label{fig:great_circles}
\end{figure}

The result is a great circle that goes through $P_{0}$, as needed,
and can be pointed in any direction by adjusting $\Phi$. Figure~\ref{fig:great_circles}a
shows two great circles going through $P_{0}$ in such a way that
their tangents are perpendicular at that point.

Using Eq.~(\ref{eq:XYZtoU},~\ref{eq:XYZtoV}) the expressions for
the great circles on a hemi-sphere in stereographic coordinates become:

\begin{multline}
u\left(\psi;\,\Phi,\,\theta_{0},\,\phi_{0}\right)=\Biggl(\cos\Phi\cos\theta_{0}\cos\phi_{0}\sin\psi-\\
-\sin\Phi\sin\phi_{0}\sin\psi+\sin\theta_{0}\cos\phi_{0}\cos\psi\Biggr)\times\\
\times\Biggl(1-\cos\Phi\sin\theta_{0}\sin\psi+\cos\theta_{0}\cos\psi\Biggr)^{-1}\label{eq:great_circle_u}
\end{multline}

\begin{multline}
v\left(\psi;\,\Phi,\,\theta_{0},\,\phi_{0}\right)=\Biggl(\cos\Phi\cos\theta_{0}\sin\phi_{0}\sin\psi+\\
+\sin\Phi\cos\phi_{0}\sin\psi+\sin\theta_{0}\sin\phi_{0}\cos\psi\Biggr)\times\\
\times\Biggl(1-\cos\Phi\sin\theta_{0}\sin\psi+\cos\theta_{0}\cos\psi\Biggr)^{-1}\label{eq:great_circle_v}
\end{multline}

Figure~\ref{fig:great_circles}b shows the great-circle curves, in
the north hemi-sphere, plotted in stereographic projection, with $u$
and $v$ as Cartesian coordinates. 

\section{Radiation from square aperture illuminated with phase gradient\label{app:illum_app}}

Here we restate the result from \cite{CollinBook1985} (Sec.~4.1)
specific to the situation at hand. For a square aperture, lying in
the xy-plane, with centre at origin, and driven with electric field
$\mathbf{E}=E_{0}\mathbf{\hat{x}}\exp\left(-i\left(\zeta_{x}x+\zeta_{y}y\right)\right)$,
the far-field radiation is given by (see Eq.~(4.24) in ref.~\cite{CollinBook1985}):
\begin{multline*}
\mathbf{E}\left(r,\,\theta,\,\phi\right)=\frac{ik_{0}4a^{2}E_{0}}{2\pi r}\cdot\exp\left(-ik_{0}r\right)\times\\
\times\frac{\sin\left(a\cdot\left[k_{0}\sin\theta\cos\phi-\zeta_{x}\right]\right)}{a\cdot\left[k_{0}\sin\theta\cos\phi-\zeta_{x}\right]}\times\frac{\sin\left(a\cdot\left[k_{0}\sin\theta\sin\phi-\zeta_{y}\right]\right)}{a\cdot\left[k_{0}\sin\theta\sin\phi-\zeta_{y}\right]}\\
\times\biggl(\boldsymbol{\hat{\theta}}\cos\phi-\boldsymbol{\hat{\phi}}\sin\phi\cos\theta\biggr)
\end{multline*}

Restating the problem in stereographic coordinates (using Eq.~(\ref{eq:SpherToU},~\ref{eq:SpherToV},~\ref{eq:ThetaHatFromUV},~\ref{eq:PhiHatFromUV})):

\begin{multline*}
\mathbf{E}\left(u,\,v,\,r\right)=\frac{ik_{0}4a^{2}E_{0}}{2\pi r}\times\exp\left(-ik_{0}r\right)\times\\
\times\frac{\sin\left(a\cdot\left[2k_{0}u/\alpha-\zeta_{x}\right]\right)}{a\cdot\left[2k_{0}u/\alpha-\zeta_{x}\right]}\times\frac{\sin\left(a\cdot\left[2k_{0}v/\alpha-\zeta_{y}\right]\right)}{a\cdot\left[2k_{0}v/\alpha-\zeta_{y}\right]}\times\\
\times\left(\left[\frac{u^{2}\left(1+u^{2}\right)+v^{2}\left(1-v^{2}\right)}{\left(u^{2}+v^{2}\right)\alpha}\right]\mathbf{\hat{u}}+\left[\frac{2uv}{\alpha}\right]\mathbf{\hat{v}}\right)
\end{multline*}

Where $\alpha=1+u^{2}+v^{2}$.
\end{document}